\documentstyle[aps,prl,floats,epsf]{revtex}

\def\Journal#1#2#3#4{{#1} {\bf #2}, #3 (#4)}

\def\NCA{Nuovo Cimento}

\def\NIMA{{Nucl. Instrum. Methods} A}

\def\PLB{{Phys. Lett.}  B}
\def\PRL{Phys. Rev. Lett.}
\def\PRD{{Phys. Rev.} D}

\begin{document}

\draft

\wideabs{

\title{Precision Measurement of Cosmic-Ray Antiproton Spectrum}
\author{
 S.\thinspace Orito,$^{1}$
 T.\thinspace Maeno,$^{1}$
 H.\thinspace Matsunaga,$^{1}$
 K.\thinspace Abe,$^{1}$
 K.\thinspace Anraku,$^{1}$
 Y.\thinspace Asaoka,$^{1}$
 M.\thinspace Fujikawa,$^{1}$
 M.\thinspace Imori,$^{1}$
 M.\thinspace Ishino,$^{7}$
 Y.\thinspace Makida,$^{2}$
 N.\thinspace Matsui,$^{1}$
 H.\thinspace Matsumoto,$^{3}$
 J.\thinspace Mitchell,$^{4}$
 T.\thinspace Mitsui,$^{3}$
 A.\thinspace Moiseev,$^{4}$
 M.\thinspace Motoki,$^{1}$
 J.\thinspace Nishimura,$^{1}$
 M.\thinspace Nozaki,$^{3}$
 J.\thinspace Ormes,$^{4}$
 T.\thinspace Saeki,$^{1}$
 T.\thinspace Sanuki,$^{1}$
 M.\thinspace Sasaki,$^{3}$
 E.\thinspace S.\thinspace Seo,$^{5}$
 Y.\thinspace Shikaze,$^{1}$
 T.\thinspace Sonoda,$^{1}$
 R.\thinspace Streitmatter,$^{4}$
 J.\thinspace Suzuki,$^{2}$
 K.\thinspace Tanaka,$^{2}$
 I.\thinspace Ueda,$^{1}$
 N.\thinspace Yajima,$^{6}$
 T.\thinspace Yamagami,$^{6}$
 A.\thinspace Yamamoto,$^{2}$
 T.\thinspace Yoshida,$^{2}$
 and
 K.\thinspace Yoshimura$^{1}$
}
\address{
$^{1}$University of Tokyo, Tokyo 113-0033, Japan\\
$^{2}$High Energy Accelerator Research Organization (KEK),
Tsukuba, Ibaraki 305-0801, Japan\\
$^{3}$Kobe University, Kobe, Hyogo 657-8501, Japan\\
$^{4}$National Aeronautics and Space Administration,
Goddard Space Flight Center, Greenbelt,MD 20771, USA\\
$^{5}$University of Maryland, College Park, MD 20742, USA\\
$^{6}$The Institute of Space and Astronautical Science (ISAS),
Sagamihara, Kanagawa 229-8510, Japan\\
$^{7}$Kyoto University, Kyoto 606-8502, Japan
}
\date{\today}
\maketitle

\begin{abstract}
The energy spectrum of cosmic-ray antiprotons ($\bar{p}$'s)
has been measured in the range 0.18 to 3.56 GeV, based on 458 $\bar{p}$'s
collected by BESS in recent solar-minimum period.
We have detected for the first time a distinctive peak at 2 GeV of $\bar{p}$'s
originating from cosmic-ray interactions with the interstellar gas.
The peak spectrum is reproduced by theoretical calculations,
implying that the propagation models are basically correct and that
different cosmic-ray species undergo a universal propagation.
Future BESS flights toward the solar maximum will help us
to study the solar modulation and the propagation in detail
and to search for primary $\bar{p}$ components.
\end{abstract}


}

\narrowtext
The origin of cosmic-ray antiprotons ($\bar{p}$'s) has attracted
much attention since their observation
was first reported by Golden {\it et al.}\cite{GOLDEN}.
Cosmic-ray $\bar{p}$'s should certainly be produced by the interaction
of Galactic high-energy cosmic-rays with the interstellar medium.
The energy spectrum of these ``secondary'' $\bar{p}$'s is expected to show
a characteristic peak around 2 GeV, with sharp decreases of the flux
below and above the peak, a generic feature which reflects
the kinematics of $\bar{p}$ production.
The secondary $\bar{p}$'s offer a unique probe \cite{GS92} of
cosmic-ray propagation and of solar modulation.
As other possible sources of cosmic-ray $\bar{p}$'s,
one can conceive novel processes,
such as annihilation of neutralino dark matter or
evaporation of primordial black holes \cite{HA75}.  The
$\bar{p}$'s from these ``primary'' sources, if they exist, are expected to be
prominent at low energies \cite{MA96}
and to exhibit large solar modulations
\cite{MI96}.
Thus they are distinguishable in principle
from the secondary $\bar{p}$ component.

The detection of the secondary peak and the search for
a possible low-energy primary $\bar{p}$ component have been difficult
to achieve, because of huge backgrounds and the
extremely small flux especially at low energies.
The first \cite{GOLDEN} and subsequent \cite{BOGO} evidence
for cosmic-ray $\bar{p}$'s  were reported at relatively high energies,
where it was not possible to positively identify
the $\bar{p}$'s with a mass measurement.
The first ``mass-identified'' and thus unambiguous detection
of cosmic-ray $\bar{p}$'s was
performed by BESS '93 \cite{YOMO}
in the low-energy region (4 events at 0.3 to 0.5 GeV),
which was followed by IMAX \cite{IMAX}
and CAPRICE \cite{CAPRICE} detections.
The BESS '95 measured
the spectrum \cite{MA98} at solar minimum, based
on 43 $\bar{p}$'s over the range 0.18 to 1.4 GeV.
We report here a new high-statistics measurement of the $\bar{p}$ spectrum
based on 458 events in the energy range from 0.18 to 3.56 GeV.

\begin{figure}[b]
\centerline{\epsfxsize=8cm \epsffile{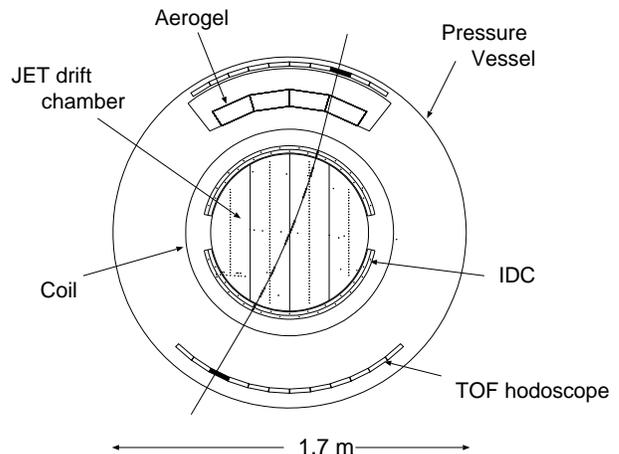}}
\caption{
Cross-sectional view of the BESS spectrometer in its 1997 configuration.
Overlayed is one of the $\bar{p}$ events.
}
\label{fig:detector}
\end{figure}

Fig.\ref{fig:detector} shows a schematic view of BESS.
It was designed \cite{OR87} and constructed \cite{DETE}
as a high-resolution spectrometer to perform
searches for rare cosmic-rays, as well as
various precision measurements.
A uniform field of 1 Tesla is produced by a thin
(4 g/cm$^2$) superconducting coil \cite{YA88}, through
which particles can pass without too many interactions.  The
magnetic-field region is filled with the tracking
volume.  This geometry results in an acceptance of
0.3 m$^2$sr, which is an order of magnitude larger than those of
previous cosmic-ray spectrometers.
The tracking
is performed by fitting up to 28 hit-points in the drift
chambers, resulting in a magnetic-rigidity ($R$) resolution of
0.5 \% at 1 GV/$c$.
The upper and lower scintillator-hodoscopes provide
two d$E$/d$x$ measurements and the
time-of-flight (TOF) of particles. The d$E$/d$x$ in the
drift chamber gas is obtained as a truncated mean of the
integrated charges of the hit-pulses. For the '97 flight, the
hodoscopes were placed at the outer-most radii, and
the timing resolution of each counter was improved to
50 psec rms,
resulting in $\beta^{-1}$ resolution of 0.008, where $\beta$ is
defined as particle velocity \cite{BETA} divided by the speed of the light.
Furthermore, a Cherenkov counter with a silica-aerogel ($n$ = 1.032) radiator
was newly installed \cite{AS98}, in order to veto
$e^{-}/\mu^{-}$ backgrounds
which gave large Cherenkov light outputs corresponding to
14.7 mean photo-electrons when crossing the aerogel.

The 1997 BESS balloon flight was carried out on July 27,
from Lynn Lake, Canada.
The scientific data were taken for 57,032 sec of live time
at altitudes ranging from 38 to 35 km
(an average residual air of 5.3 g/cm$^2$) and cut-off rigidity
ranging from 0.3 to 0.5 GV/$c$.
The first-level trigger
was provided by a coincidence between the top and the bottom
scintillators, with the threshold set at 1/3 of the pulse height
from minimum ionizing particles.  The second-level trigger, which
utilized the hit-patterns of the hodoscopes and
the inner drift chambers (IDC), first rejected unambiguous null- and
multi-track events and made a rough rigidity-determination
to select negatively-charged particles predominantly.
In addition, one of every 60
first-level triggers was recorded,
in order to build a sample of unbiased triggers.

\begin{figure}[b]
\centerline{\epsfxsize=8cm \epsffile{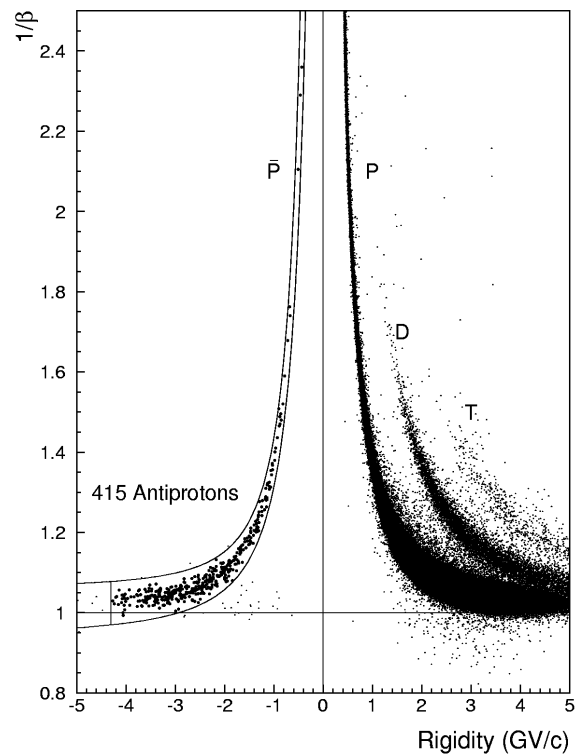}}
\caption{
The identification of $\bar{p}$ events.
The solid lines define the $\bar{p}$ mass band used for
the spectrum measurement.
}
\label{fig:id97}
\end{figure}

The off-line analysis \cite{MA98} selects events
with a single track fully contained
in the fiducial region of the tracking volume
with acceptable track qualities.
The three d$E$/d$x$ measurements are loosely required as function of $R$
to be compatible with proton or $\bar{p}$.
The combined efficiency of these off-line selections
is 83 -- 88 \% for $R$ from 0.5 to 4 GV/$c$.
These simple and highly-efficient selections are sufficient for a very
clean detection of $\bar{p}$'s in the low-velocity ($\beta < 0.9$) region.
At higher-velocities, the $e^{-}/\mu^{-}$ background
starts to contaminate the $\bar{p}$ band, where
we require the Cherenkov veto; i.e.,
1) the particle trajectory to cross the fiducial volume of the aerogel,
and 2) the Cherenkov output
to be less than 0.09 of the mean output from $e^{-}$.
This cut reduces the acceptance by 20 \%,
but rejects $e^{-}/\mu^{-}$ backgrounds by a factor of 6000, while
keeping 93 \% efficiency for protons and $\bar{p}$'s
which cross the aerogel with rigidity below the threshold
(3.8 GV/$c$).
Fig.\ref{fig:id97} shows the $\beta^{-1}$ versus $R$ plot
for the surviving events.  We see a clean narrow band of 415 $\bar{p}$'s
at the exact mirror position of the protons.  The $\bar{p}$
sample is thus mass-identified and background-free, as the cleanness
of the band demonstrates and various background studies show.
In particular, backgrounds of albedo and of mis-measured
positive-rigidity particles are totally excluded
by the excellent $\beta^{-1}$ and $R^{-1}$ resolutions.
To check against the ``re-entrant albedo''
background, we confirmed that the trajectories of all
$\bar{p}$'s can be traced numerically through the Earth's
geomagnetic field back to the outside of the geomagnetic sphere.

We obtain the $\bar{p}$ fluxes at the top of the atmosphere
(TOA) in the following way:
The geometrical acceptance of the spectrometer is
calculated both analytically and by two independent Monte Carlo methods.
The live data-taking time is directly measured by two independent scaler
systems gated by the ``ready'' gate which
controls the first-level trigger.
The efficiencies of the second-level trigger and of the
off-line selections are determined by using the unbiased
trigger sample. The TOA energy of each event is calculated
by tracing back the particle through the detector material and the air.
The interaction loss of the $\bar{p}$'s is evaluated
by applying the same selections to the Monte Carlo events
generated by {\sc geant/gheisha}, which incorporates \cite{MATSU}
detailed material distribution
and correct $\bar{p}$-nuclei cross sections.  We subtract
the expected number \cite{MITSUI} of atmospheric $\bar{p}$'s, produced by the
collisions of cosmic-rays in the air.  The subtraction amounts to
$9\pm2$ \%, $15\pm3$ \% and $19\pm5$ \%, respectively, at 0.25, 0.7
and 2 GeV, where the errors correspond to the maximum difference
among three recent calculations \cite{MITSUI,PF96,ST96}
which agree to each other.
Proton fluxes are obtained in a similar way. Atmospheric protons are
subtracted by following Papini \cite{PAPI}.

Table \ref{tab:pbsummary} contains the resultant BESS '97 $\bar{p}$ fluxes
and $\bar{p}/p$ flux ratios at TOA.
The first and the second errors
represent the statistical \cite{FEL98} and systematic errors, respectively.
We checked that the central values of the fluxes are stable against
various trial changes of the selection criteria,
including uniform application of the
Cherenkov veto also to the low $\beta$ region.
The dominant systematic errors at high and low energies, respectively,
are uncertainties in the atmospheric $\bar{p}$ calculations and
in the $\bar{p}$ interaction losses
to which we attribute $\pm$15 \% relative error.
As shown in Table \ref{tab:pbsummary},
the BESS '97 fluxes are consistent \cite{NOTE1} with the '95 fluxes
in the overlapping low-energy range (0.2 to 1.4 GeV).  The solar
activities at the time of the two flights were both close to
the minimum as shown by world neutron monitors and by the low-energy
proton spectra \cite{SAN98} measured by BESS.

\widetext
\begin{table}[t]
\caption{
Antiproton fluxes
(in $\times 10^{-2}$ m$^{-2}$s$^{-1}$sr$^{-1}$GeV$^{-1}$)
and $\bar{p}/p$ ratios (in $\times 10^{-5}$) at TOA.
T (in GeV) define the kinetic energy bins.
${\rm{N}}_{\bar{p}}$ and $\overline{\rm{T}}_{\bar{p}}$,
respectively, are the number of observed antiprotons
and their mean kinetic energy in each bin.
The eighth bin of BESS '95 flux actually covers the energy region
from 1.28 to 1.40 GeV.
}
\label{tab:pbsummary}
\renewcommand{\arraystretch}{1.25}
\begin{tabular}{c|cccc|ccc|ccc}

 & BESS
 & '97
 & 
 & 
 & BESS
 & '95
 & 
 & BESS
 & '97+'95
 & \\
T (GeV)
 & ${\rm{N}}_{\bar{p}}$
 & $\overline{\rm{T}}_{\bar{p}}$
 & $\bar{p}$ flux
 & $\bar{p}/p$ ratio
 & ${\rm{N}}_{\bar{p}}$
 & $\overline{\rm{T}}_{\bar{p}}$
 & $\bar{p}$ flux
 & $\overline{\rm{T}}_{\bar{p}}$
 & $\bar{p}$ flux
 & $\bar{p}/p$ ratio\\
\tableline
0.18 - 0.28
 &     4
 &   0.21
 & $0.74^{+0.58\,+0.12}_{-0.34\,-0.12} $
 & $0.44^{+0.34\,+0.08}_{-0.20\,-0.08} $
 &     3
 &   0.24
 & $1.75^{+1.41\,+0.37}_{-1.13\,-0.37} $
 &   0.22
 & $1.00^{+0.51\,+0.18}_{-0.42\,-0.18} $
 & $0.51^{+0.31\,+0.08}_{-0.19\,-0.08} $ \\
0.28 - 0.40
 &     9
 &   0.35
 & $1.05^{+0.51\,+0.12}_{-0.36\,-0.12} $
 & $0.52^{+0.25\,+0.08}_{-0.18\,-0.08} $
 &     3
 &   0.34
 & $1.00^{+0.86\,+0.14}_{-0.66\,-0.14} $
 &   0.35
 & $1.04^{+0.43\,+0.12}_{-0.31\,-0.12} $
 & $0.52^{+0.22\,+0.06}_{-0.16\,-0.06} $ \\
0.40 - 0.56
 &    16
 &   0.49
 & $1.23^{+0.45\,+0.13}_{-0.34\,-0.13} $
 & $0.67^{+0.24\,+0.10}_{-0.18\,-0.10} $
 &     6
 &   0.49
 & $1.40^{+0.87\,+0.17}_{-0.58\,-0.17} $
 &   0.49
 & $1.27^{+0.37\,+0.14}_{-0.32\,-0.14} $
 & $0.70^{+0.22\,+0.08}_{-0.16\,-0.08} $ \\
0.56 - 0.78
 &    31
 &   0.66
 & $1.63^{+0.41\,+0.16}_{-0.37\,-0.16} $
 & $1.01^{+0.26\,+0.14}_{-0.23\,-0.14} $
 &     8
 &   0.67
 & $1.29^{+0.66\,+0.14}_{-0.54\,-0.14} $
 &   0.66
 & $1.54^{+0.33\,+0.16}_{-0.30\,-0.16} $
 & $0.97^{+0.22\,+0.10}_{-0.19\,-0.10} $ \\
0.78 - 0.92
 &    19
 &   0.85
 & $1.41^{+0.48\,+0.14}_{-0.42\,-0.14} $
 & $1.11^{+0.38\,+0.16}_{-0.33\,-0.16} $
 &     6
 &   0.83
 & $1.57^{+1.07\,+0.17}_{-0.71\,-0.17} $
 &   0.85
 & $1.44^{+0.44\,+0.15}_{-0.36\,-0.15} $
 & $1.15^{+0.35\,+0.12}_{-0.29\,-0.12} $ \\
0.92 - 1.08
 &    16
 &   1.01
 & $0.83^{+0.42\,+0.10}_{-0.32\,-0.10} $
 & $0.78^{+0.39\,+0.12}_{-0.30\,-0.12} $
 &     5
 &   0.99
 & $1.05^{+0.84\,+0.12}_{-0.65\,-0.12} $
 &   1.01
 & $0.87^{+0.36\,+0.10}_{-0.32\,-0.10} $
 & $0.82^{+0.35\,+0.09}_{-0.27\,-0.09} $ \\
1.08 - 1.28
 &    32
 &   1.19
 & $1.68^{+0.46\,+0.15}_{-0.41\,-0.15} $
 & $1.86^{+0.50\,+0.25}_{-0.46\,-0.25} $
 &     7
 &   1.18
 & $1.60^{+0.99\,+0.16}_{-0.82\,-0.16} $
 &   1.19
 & $1.65^{+0.40\,+0.15}_{-0.36\,-0.15} $
 & $1.85^{+0.46\,+0.18}_{-0.41\,-0.18} $ \\
1.28 - 1.52
 &    43
 &   1.40
 & $2.18^{+0.49\,+0.19}_{-0.44\,-0.19} $
 & $2.89^{+0.65\,+0.38}_{-0.59\,-0.38} $
 &     5
 &   1.33
 & $1.87^{+1.35\,+0.18}_{-1.08\,-0.18} $
 &   1.39
 & $2.13^{+0.43\,+0.19}_{-0.39\,-0.19} $
 & $2.82^{+0.61\,+0.25}_{-0.54\,-0.25} $ \\
1.52 - 1.80
 &    51
 &   1.65
 & $2.45^{+0.48\,+0.24}_{-0.44\,-0.24} $
 & $4.22^{+0.83\,+0.59}_{-0.76\,-0.59} $
 &   -
 &   -
 &   -
 &   1.65
 & $2.45^{+0.48\,+0.24}_{-0.44\,-0.24} $
 & $4.22^{+0.83\,+0.59}_{-0.76\,-0.59} $ \\
1.80 - 2.12
 &    51
 &   1.96
 & $2.27^{+0.45\,+0.24}_{-0.42\,-0.24} $
 & $4.90^{+0.98\,+0.71}_{-0.90\,-0.71} $
 &   -
 &   -
 &   -
 &   1.96
 & $2.27^{+0.45\,+0.24}_{-0.42\,-0.24} $
 & $4.90^{+0.98\,+0.71}_{-0.90\,-0.71} $ \\
2.12 - 2.52
 &    64
 &   2.31
 & $2.40^{+0.42\,+0.21}_{-0.37\,-0.21} $
 & $6.74^{+1.19\,+0.89}_{-1.03\,-0.89} $
 &   -
 &   -
 &   -
 &   2.31
 & $2.40^{+0.42\,+0.21}_{-0.37\,-0.21} $
 & $6.74^{+1.19\,+0.89}_{-1.03\,-0.89} $ \\
2.52 - 3.00
 &    56
 &   2.72
 & $2.02^{+0.40\,+0.18}_{-0.35\,-0.18} $
 & $7.80^{+1.54\,+1.05}_{-1.34\,-1.05} $
 &   -
 &   -
 &   -
 &   2.72
 & $2.02^{+0.40\,+0.18}_{-0.35\,-0.18} $
 & $7.80^{+1.54\,+1.05}_{-1.34\,-1.05} $ \\
3.00 - 3.56
 &    23
 &   3.25
 & $1.65^{+0.56\,+0.20}_{-0.44\,-0.20} $
 & $7.63^{+2.59\,+1.19}_{-2.04\,-1.19} $
 &   -
 &   -
 &   -
 &   3.25
 & $1.65^{+0.56\,+0.20}_{-0.44\,-0.20} $
 & $7.63^{+2.59\,+1.19}_{-2.04\,-1.19} $ \\
\end{tabular}
\end{table}

\renewcommand{\arraystretch}{1}
\narrowtext

\begin{figure}[b]
\centerline{\epsfxsize=8.5cm \epsffile{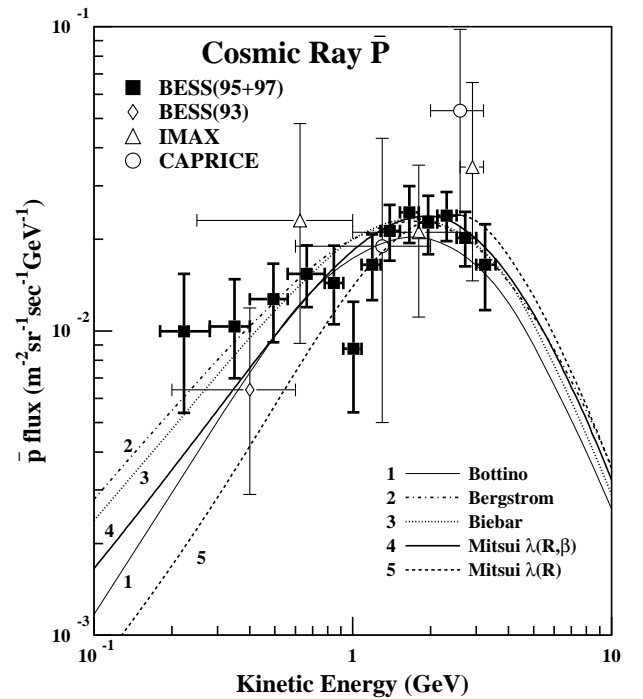}}
\caption{
BESS '95+'97 (solar minimum) antiproton fluxes at the top of the atmosphere
together with previous data.
The error-bars represent the quadratic
sums of the statistical and systematic errors.
The curves are recent calculations of the secondary
$\bar{p}$ spectra for the solar minimum period.
}
\label{fig:pbflux}
\end{figure}

Shown in Fig.\ref{fig:pbflux} is the combined BESS ('95+'97) spectrum,
in which we detect for the first time a distinctive peak
at 2 GeV of secondary $\bar{p}$, which clearly is the dominant
component of the cosmic-ray $\bar{p}$'s.

The measured secondary $\bar{p}$ spectrum provides crucial tests
of models of propagation and solar modulation,
since one has {\it a priori} knowledge of the input source spectrum
for the secondary $\bar{p}$, which can be calculated
by combining the measured proton and helium spectra with
the accelerator data\cite{TN83} on the $\bar{p}$ production.
The distinct peak structure of the $\bar{p}$ spectrum also has
clear advantages in these tests
over the monotonic (and unknown) source spectra of other cosmic-rays.

The curves shown in Fig.\ref{fig:pbflux} are recent theoretical
calculations for the secondary $\bar{p}$
in diffusion model \cite{BOTT,BERG}
and leaky box model \cite{BIEB,MITS99},
in which the propagation parameters (diffusion-coefficient
or escape length) are deduced
by fitting various data on cosmic-ray nuclei such as Boron/Carbon ratio,
under the assumption that
the different cosmic-ray species (nuclei, proton and $\bar{p}$)
undergo a universal propagation process.
These calculations
\mbox{ } \hspace*{8.5cm} \vspace*{8.8cm} \mbox{ }
use as essential inputs recently
measured proton spectra \cite{SAN98,MENN,BOEZ},
which are significantly
(by a factor 1.4 to 1.6) lower than
previous data in the energy range (10 to 50 GeV)
relevant to the $\bar{p}$ production.

\begin{figure}[t]
\centerline{\epsfxsize=8.5cm \epsffile{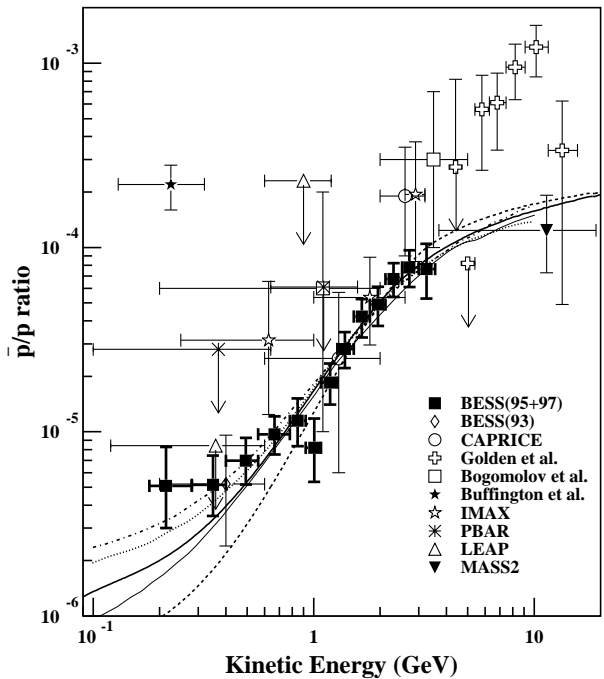}}
\caption{
BESS '95+'97 $\bar{p}/p$ flux ratios with previous data [9]
and the theoretical calculations.
}
\label{fig:ppratio}
\end{figure}

These calculations reproduce
our spectrum at the peak region remarkably well
within their $\pm$15 \% estimated accuracy \cite{MITS99}.
This implies that the propagation models \cite{MODELS} are basically correct
and that different cosmic-ray species
undergo a universal propagation process.

At low energies, the calculations predict somewhat diverse spectra
reflecting various uncertainties \cite{PROP},
which presently make it difficult to draw any conclusion on possible
admixture of primary $\bar{p}$ component.
As noted in Ref.\cite{MI96}, the rapid increase of
solar activity toward the year 2001 will drastically suppress the primary
$\bar{p}$ component such as from the primordial black holes \cite{HA75},
while changing
the shape of the secondary $\bar{p}$ only modestly \cite{BIEB}.
This will help us to separate out the ``primary'' and ``secondary''
components at low energies.
Future annual BESS flights will thus be important
to search for primary $\bar{p}$ component,
to study solar modulation,
and to investigate further details of the propagation.

Since most previous data were presented
in the form of $\bar{p}/p$ flux ratios \cite{CAPRICE},
a compilation is made in Fig.\ref{fig:ppratio},
which shows again the unprecedented accuracy of our measurement.

Sincere thanks are given to NASA and NSBF for the balloon launch.  The
analysis was performed by using the computing facilities
at ICEPP, Univ. of Tokyo.  This experiment was supported by
Grant-in-Aid from Monbusho in Japan and by NASA in the USA.


\begin{references}
\bibitem{BETA}
$\beta$ is defined to be positive for down-going particles.
\bibitem{NOTE1}
Our '97 data have a factor 2.7 times more statistics than '95 data
in the overlapping low-energy region.
This is realized by combination of longer flight-time,
less dead-time and higher trigger- and selection-efficiencies.
\bibitem{PROP}
The main differences among the calculations are the treatment
of the tertiary $\bar{p}$ interactions, nuclear models to simulate
proton-nucleus and nucleus-proton interactions, the $R$ and $\beta$
dependences of the propagation parameter, and the solar modulations.
\end{references}
\end{document}